\documentclass[a4paper]{jpconf}
\usepackage{citesort}
\usepackage{graphicx}% Include figure files
\usepackage{xcolor}
\usepackage{wrapfig}

\begin{document}
\nocite{*}

\title{Quantum self-organization and nuclear collectivities}

%\noindent \qquad \\[-6pt] \qquad Version 2.0\\\qquad December 21, 2006

\author{T~Otsuka}
\address{Department of Physics and Center for Nuclear Study, University of Tokyo, 7-3-1 Hongo, Bunkyo, Tokyo 113-0033 Tokyo, Japan}
\address{RIKEN Nishina Center, 2-1 Hirosawa, Wako, Saitama 351-0198, Japan}
\address{Instituut voor Kern- en Stralingsfysica, K. U. Leuven, Celestijnenlaan 200D, B-3001 Leuven, Belgium}

\author{Y~Tsunoda, T~Togashi, N~Shimizu}
\address{Center for Nuclear Study, University of Tokyo, 7-3-1 Hongo, Bunkyo, Tokyo 113-0033 Tokyo, Japan}

\author{T~Abe}
\address{Department of Physics, University of Tokyo, 7-3-1 Hongo, Bunkyo, Tokyo 113-0033 Tokyo, Japan}

\ead{otsuka@phys.s.u-tokyo.ac.jp}

\begin{abstract}
The quantum self-organization is introduced as one of the major underlying mechanisms of the quantum many-body systems.  In the case of atomic nuclei as an example, two types of the motion of nucleons, single-particle states and collective modes, dominate the structure of the nucleus.
%, while they are considered to compete against  each other.  
The outcome of the collective mode is determined basically by the balance
between the effect of the mode-driving force ({\it e.g.}, quadrupole force for the ellipsoidal deformation) and 
the resistance power against it.  The single-particle energies are one of the sources to produce such resistance power: a coherent collective motion is more hindered by larger gaps between relevant single particle states.  
Thus, the single-particle state and the collective mode are ``enemies" each other.  However, 
the nuclear forces are demonstrated to be rich enough so as to enhance relevant collective mode by reducing the resistance power by changing single-particle energies for each eigenstate through monopole interactions.  This will be verified with the concrete example taken from Zr isotopes.  Thus, when the quantum self-organization occurs, single-particle energies can be self-organized, being enhanced by (i) two quantum liquids, e.g., protons and neutrons, (ii) two major force components, e.g., quadrupole interaction (to drive collective mode) and monopole interaction (to control resistance). In other words, atomic nuclei are not necessarily like simple rigid vases containing almost free nucleons, in contrast to the na\"ive Fermi liquid picture.  Type II shell evolution is considered to be a simple visible case involving excitations across a (sub)magic gap. The quantum self-organization becomes more important in heavier nuclei where the number of active orbits and the number of active nucleons are larger. The quantum self-organization is a general phenomenon, and is expected to be found in other quantum systems.

\end{abstract}

\section{Introduction}
The underlying mechanisms of the quantum many-body structure of atomic nuclei have been studied over decades as one of the most important objectives of nuclear physics.   
It has then been understood commonly that there are %\textcolor{red}{
two types of dominant motion of nucleons in the atomic nucleus%}
: single-particle states and collective modes.   
Regarding the single-particle states, Mayer and Jensen introduced the shell 
structure and associated magic numbers \cite{mayer1949,haxel1949,mayer_jensen_book}.  
The nuclear shell model developed on these concepts has been extremely successful in the description of 
the structure of many nuclei  (see for example \cite{talmi_book,heyde_book,rmp_SM}).    
The collective modes include various cases.  Among them, the deformation of the nuclear shape has been studied
since Rainwater \cite{rainwater1950}, and Bohr and Mottelson \cite{bohr1952,bohr_mottelson1953}.  The nuclear shapes have been one of the major focuses of the
nuclear structure physics, including spherical, vibrational and rotational ones \cite{bohr_mottelson_book2}.
The relation between the single-particle states and the collective modes has naturally become of much interest, as described in \cite{bohr_mottelson_book2} as
``the problem of reconciling the simultaneous occurrence of single-particle and collective degrees of freedom 
and exploring the variety of phenomena that arise from their interplay".

The atomic nucleus is a many-body quantum system comprised of protons and neutrons, which   
is often considered to be a Landau's Fermi liquid.  
In a somewhat simplified expression of this picture, protons and neutrons of a nucleus are in a mean 
potential which is like a rigid ``vase'', and those nucleons are like free particles moving in this vase, interacting weakly among themselves.   
The single-particle energies (SPE) of such a system exhibit the shell structure, and are split in general.   If the splitting is large enough, the many-body structure is dominated by the SPEs: nucleons occupy the lowest single particle states in the ground state, the 
next lowest configuration gives us the first excited state, and so forth.
The correlations due to the interaction between nucleons may contribute, but their effects are suppressed by the SPE splittings more or less.   If %\textcolor{red}{
the energy gain from such correlations %}
 overcomes the relevant SPE splittings, a collective mode dominates the structure of the ground and low-lying states.
Although the understanding of the relation between the single-particle states and the collective modes has been pursued in many ways, it seems to remain an open problem.
For instance, G.E. Brown had kept, throughout his life, the question, ghow single particle states can coexist with collective modesh as quoted from ``Fermi liquid theory: A brief survey in memory of Gerald E. Brown'' in \cite{schaefer2014}.   
We shall present a novel mechanism which is closely related to this problem.    

\section{Nuclear shapes and quantum phase transition}
\begin{figure} [h]
\hspace{1cm}
\includegraphics[width=5cm] {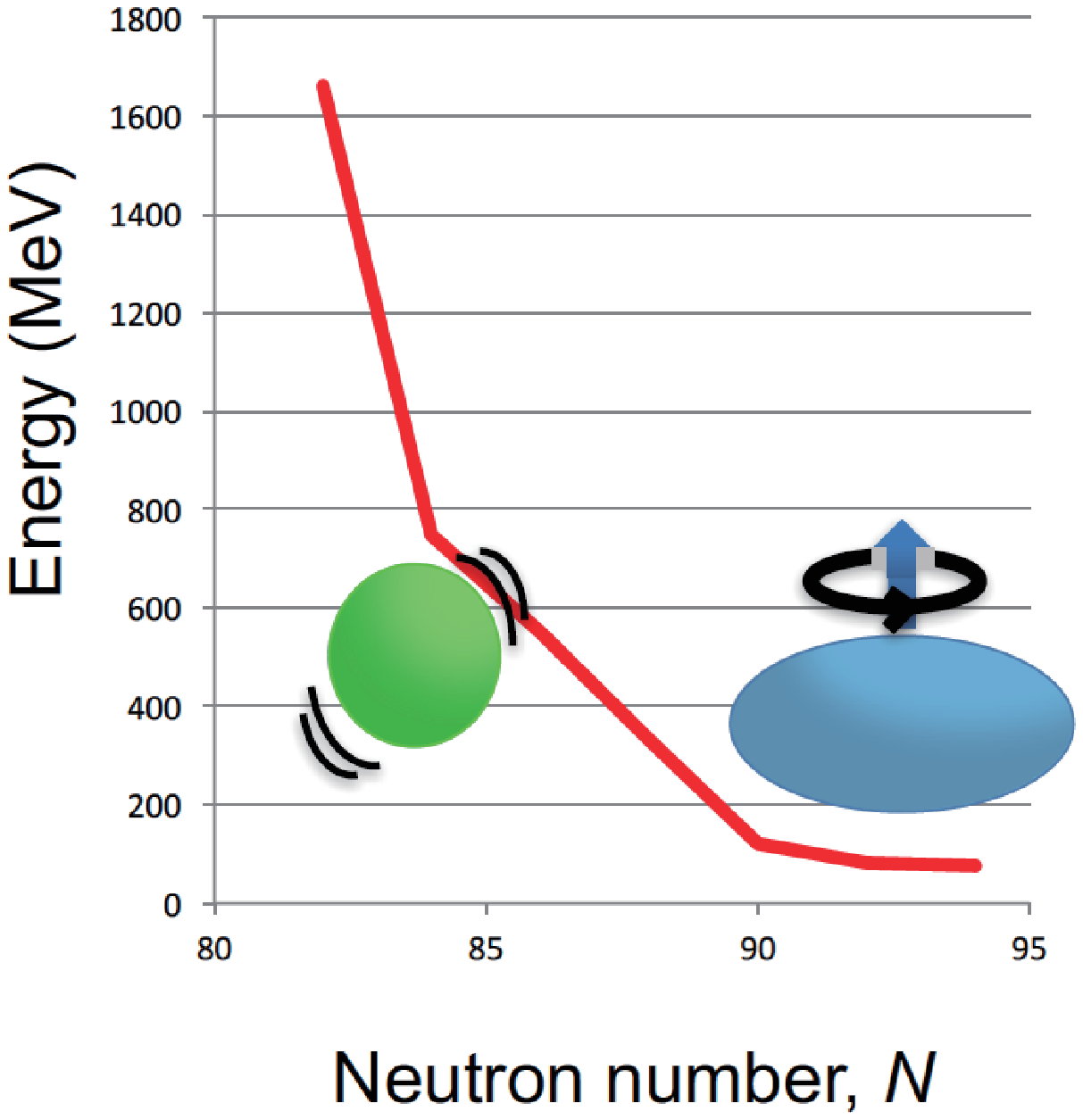}
\hspace{1cm}
\includegraphics[width=7cm] {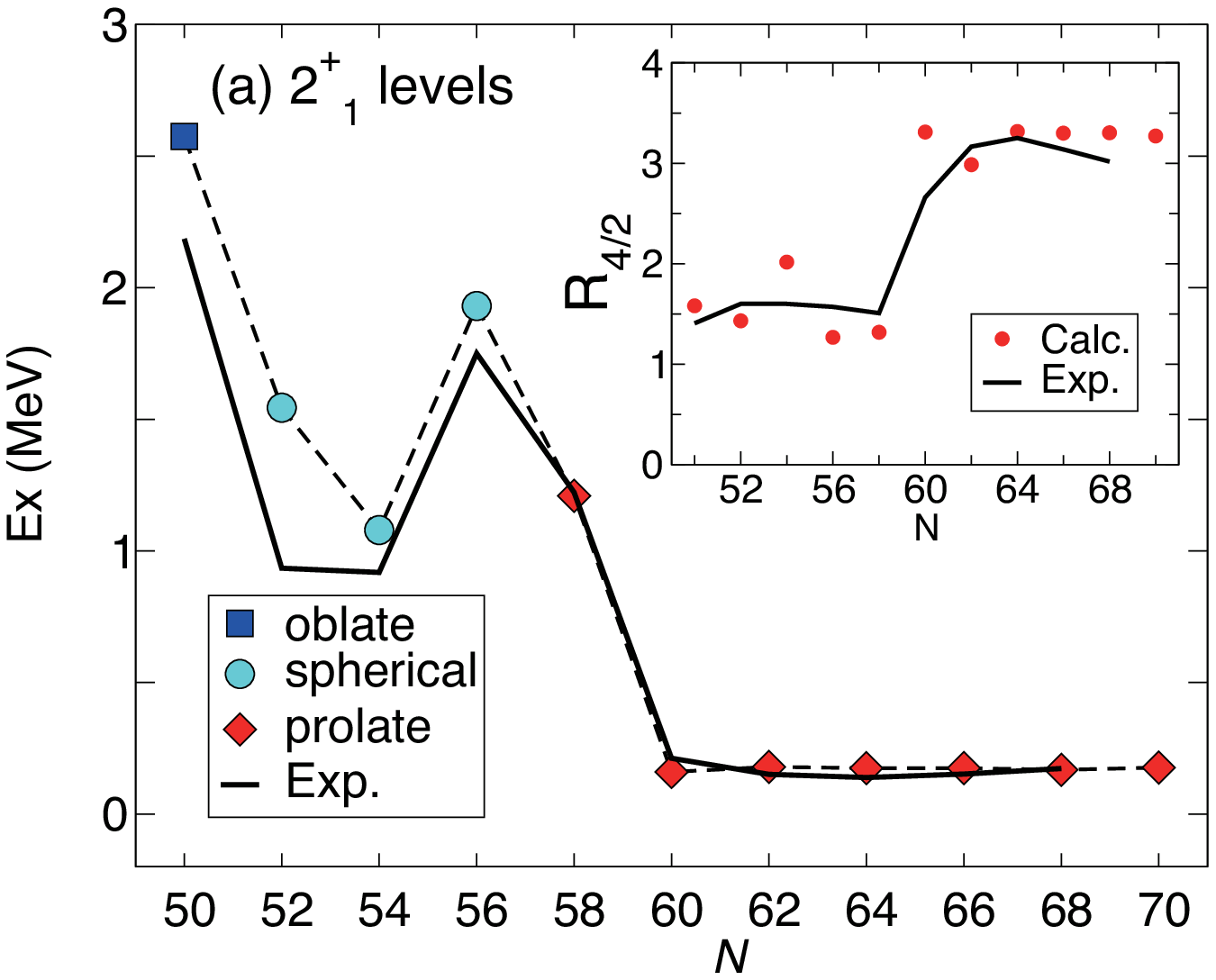}
\caption {Systemtic changes of the 2$^+_1$ level in (left) Sm and (right) Zr isotopes, 
as functions of $N$. 
Data taken from \cite{nudat2} for Sm.   The right panel is based on \cite{togashi2016}.} 
\label{fig:2+}
\end{figure}

We shall focus on the quadrupole deformation of the nuclear shape in this paper, while the scope is general.  Figure~\ref{fig:2+} exhibits the excitation energy of 2$^+_1$ state, or the 2$^+_1$ level, for Sm and Zr isotopes as a function of the neutron number, $N$.  In the Sm chain, the 2$^+$ level comes down rather gradually, similarly to many other isotopic chains.  As shown in Fig.~\ref{fig:2+}, a higher 2$^+$ level corresponds to a spherical shape and its surface oscillation, while a lower 2$^+$ level implies an ellipsoidal deformed shape and its rotation.  On the other hand,  in the Zr chain, the 2$^+$ level drops down abruptly from $N$=58 to 60.  Due to the abrupt change, this phenomenon can be called a quantum phase transition \cite{togashi2016}.  
Likewise, the ground-state structure of the Zr isotopes is changed drastically between $N$=58 and 60, also from 
the spherical to the strongly deformed shapes.   The Monte Carlo Shell Model (MCSM) describes both situations including the abrupt change with the same Hamiltonian  \cite{togashi2016,kremer2016}.  

\section{Quantum Phase Transition in Zr isotopes}
\begin{figure} [h]
\includegraphics[width=5.6cm] {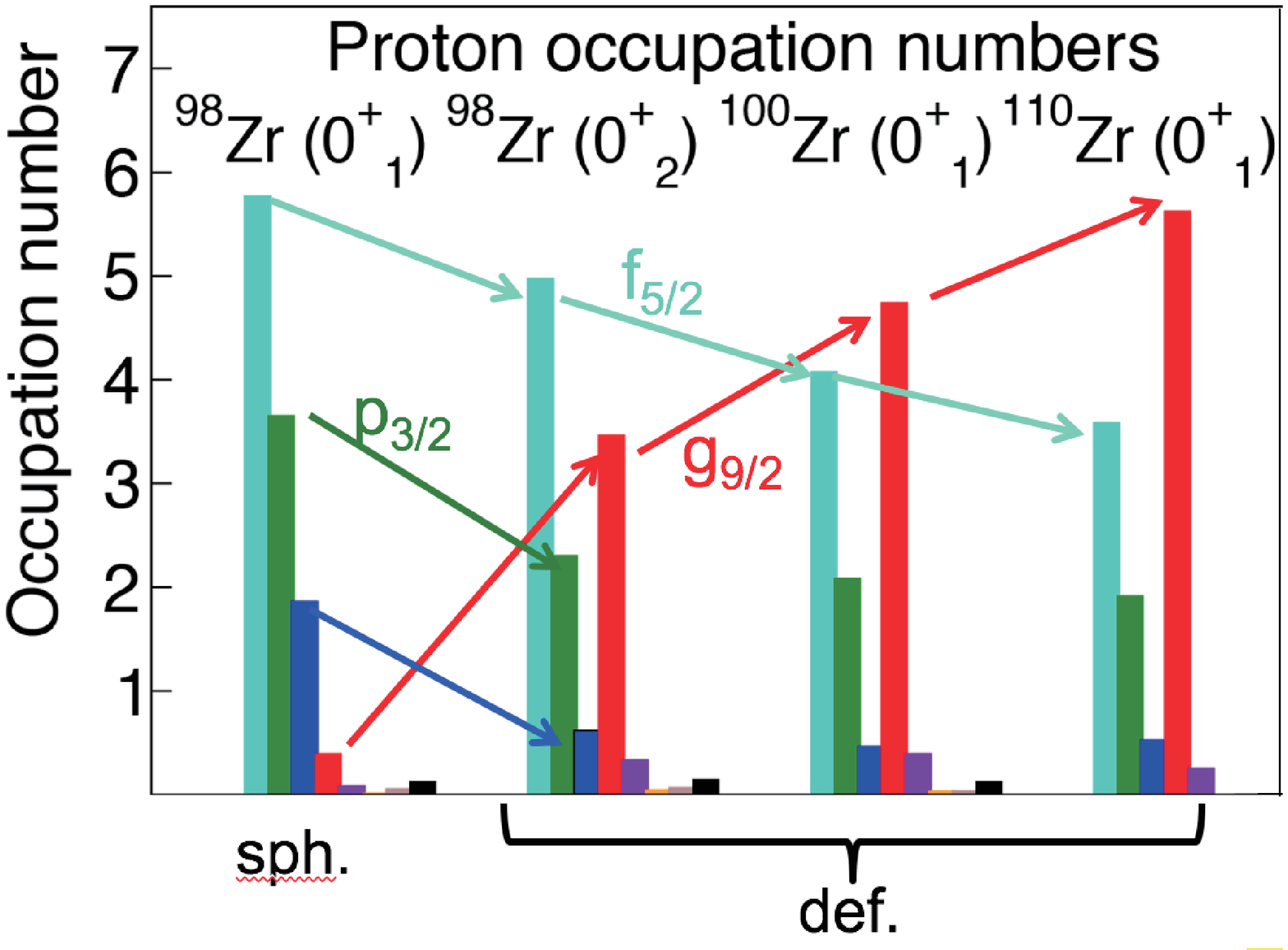}
%\hspace{0.05cm}
\includegraphics[width=3.3cm] {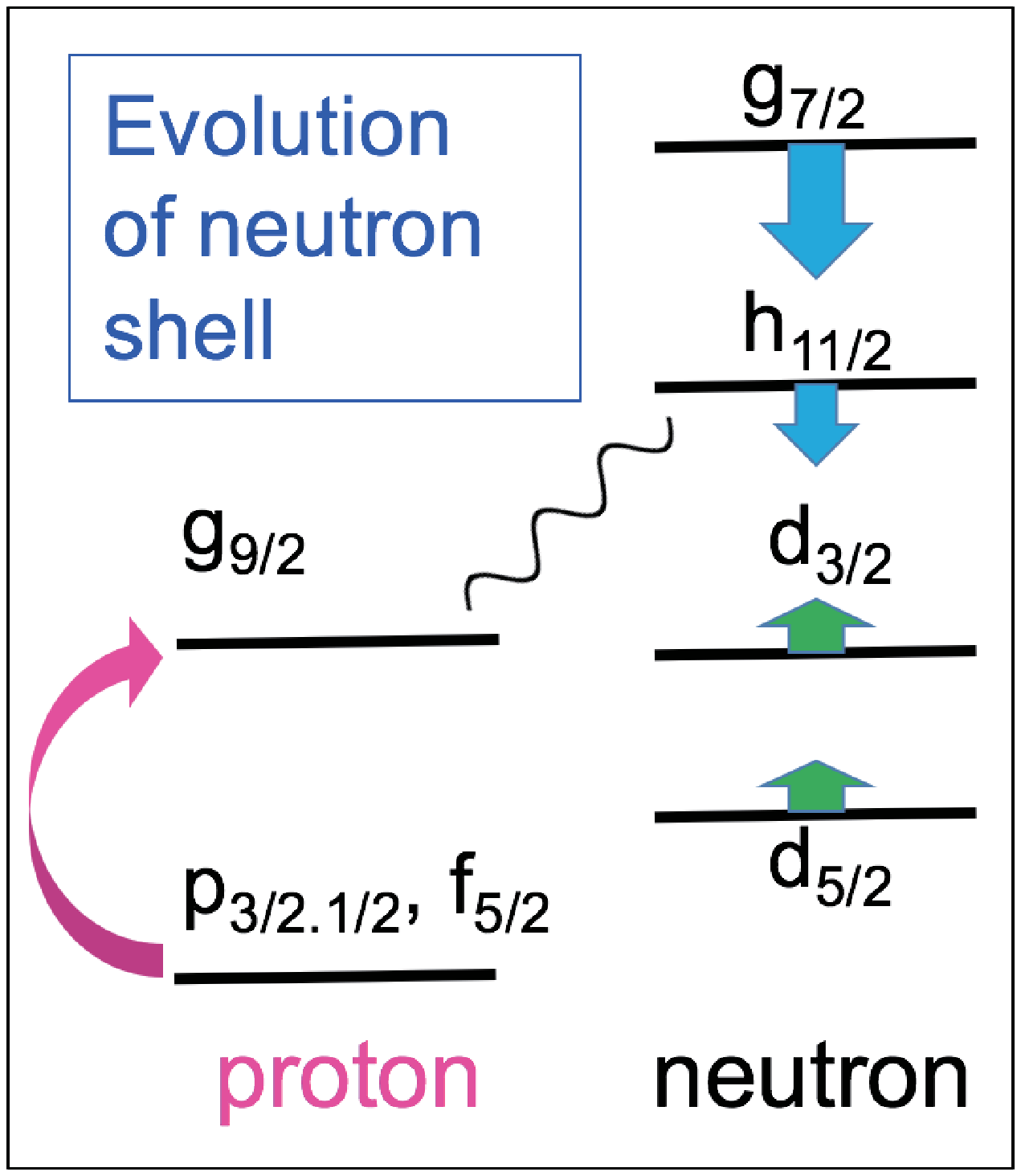}
%\hspace{0.05cm}
\includegraphics[width=6.4cm] {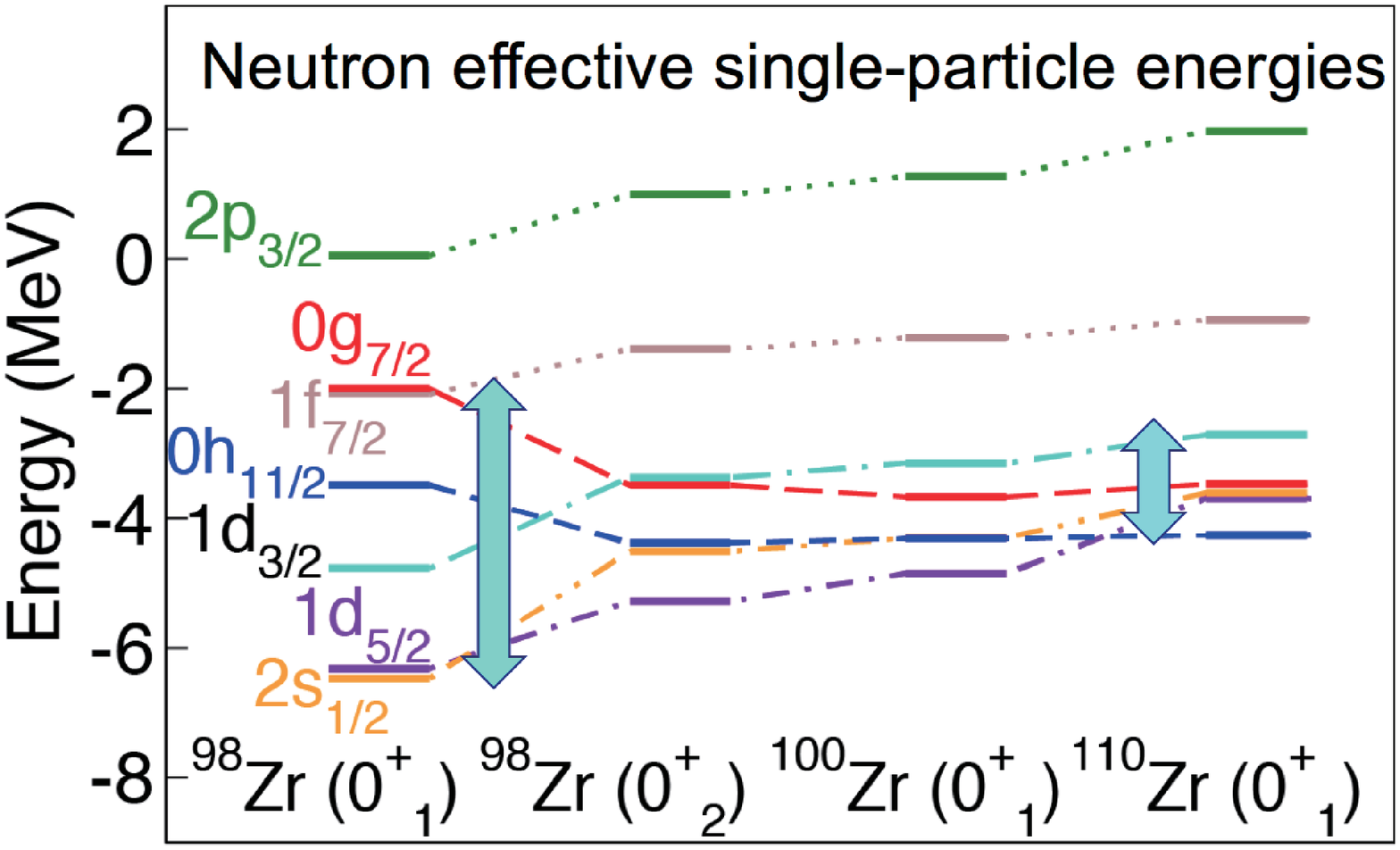}
\caption {(left) Occupation numbers of proton orbits of Zr isotopes.
(middle) Schematic illustration of the changes of neutron (effective) single-particle energies in Zr isotopes. 
(right) Actual values of neutron (effective) single-particle energies obtained in the calculation of  \cite{togashi2016}.
Left and right panels are based on \cite{togashi2016}.} 
\label{fig:Zr}
\end{figure}

We shall look into the structure changes in Zr isotopes.  The left panel of Fig.~\ref{fig:Zr} shows 
the occupation numbers of proton orbits for some states.  The g$_{9/2}$ orbit is almost empty in the 
0$^+_1$ state of $^{98}$Zr, whereas it is occupied by about 3.5 protons in its 0$^+_2$ state.
Note that this 0$^+_1$ (0$^+_2$) state is spherical (deformed).
Such changes, including the numbers of proton holes in the $pf$ orbits, result in substantial shifts of
the neutron (effective) SPEs as shown schematically in the middle panel of Fig.~\ref{fig:Zr}.
The proton-neutron monopole interaction (wavy line in the figure) generates those shifts.
The right panel depicts the actual neutron effective SPEs.  One notices a substantial change in SPE for
the different states considered.
One sees that the spacing between the d$_{5/2}$ and g$_{7/2}$ orbits is nearly 5 MeV for the 0$^+_1$ state of $^{98}$Zr, 
whereas it is reduced to about 2 MeV in 0$^+_2$ state.  Such a reduced splitting is found
also in the 0$^+_1$ state of $^{100}$Zr which is also strongly deformed.

We now discuss why the SPEs are so different between spherical and deformed states.
First, the nuclear deformation at low excitation energy is a Jahn-Teller effect \cite{jahn1937}, which 
means that the collective motion causing the deformation occurs as a consequence of coherent contributions from 
some relevant orbits near the Fermi energy.
For such coherent effects, %\textcolor{red}{
larger splittings of SPEs weaken the coherence, leading to less collectivity. %}
  On the other hand, the monopole 
interaction can change the effective SPEs depending on the occupancy of the other nucleons.
If the monopole interaction were uniform, no configuration dependence would appear, and this change should
be absent.   The tensor-force component of the nuclear force makes the monopole interaction 
attractive or repulsive, depending on the combination of the orbits \cite{otsuka2005,otsuka2010}.  This is certainly 
against the uniformity, and its effect can be crucial.  The central-force component changes its magnitude also depending on the combination of the orbits
mainly due to varying overlaps of radial wave functions of single-particle states.  Thus, the monopole interaction is indeed far from being
uniform, and the selection of favored configurations can move the SPEs of relevant orbits substantially.
If relevant SPEs can be made closer to one another, it helps the deformation.   
We shall formulate this novel mechanism in the next section.  
   
\section{Quantum self-organization and its appearance in Sm and Hg isotopes}
The nuclear deformation is determined by the balance between the effect of the collective-mode driving force  
and the resistance power against this collective mode.   This property is expressed schematically as
\begin{equation}
{\rm deformation} = \frac{\,\,\,\,{\rm quadrupole \,\, force}\,\,\,\,}{\,\,\,\,{\rm resistance \,\,power}\,\,\,\,}.
\label{eq:def}
\end{equation}
The collective-mode driving force is the quadrupole (or quadrupole-quadrupole) 
interaction in the case of the ellipsoidal shape.  This interaction is one of the major components of the proton-neutron realistic force.   A typical example of the resistance power is the pairing interaction, which tends to make the shape more   
spherical because all time-reversal pairs are equally favored.  Keeping the pairing interaction aside, we shall consider another source of the resistance power.   That is the SPEs.   

We here propose a novel mechanism, {\bf Quantum Self-Organization}.  
This mechanism means the following property:
Atomic nuclei can ``organize'' their single-particle energies by taking particular configurations
of protons and neutrons, optimized for each eigenstate, thanks to orbit-dependences of monopole components of nuclear forces ({\it e.g.}, tensor force).
This results in an enhancement of Jahn-Teller effect, {\it i.e.}, an enhancement of the collective mode.   
The deformation and quantum self-organization can be linked in a non-linear way 
%\textcolor{red}{
with a positive feedback%}
: once some nucleons are 
excited to certain orbits, the SPEs are shifted in favor of a larger deformation.   A  larger deformation can 
promote such excitations with more nucleons.  This cycle continues until a self-consistency is achieved, whereas 
intermediate situations are skipped.
In many cases, massive excitations are involved, and the particle-hole hierarchy is broken, 
for instance, a 6p-6h deformed state comes right after a 2p-2h near-spherical state, skipping 4p-4h state \cite{otsuka2016}.

The property shown in eq.~(\ref{eq:def}) is somewhat analogous to the relation,  
\begin{equation}
{\rm electric\,\,current} = \frac{\,\,\,\,{\rm voltage}\,\,\,\,}{\,\,\,\,{\rm resistance}\,\,\,\,}, 
\label{eq:current}
\end{equation}
where the electric current, voltage and resistance mean the usual quantities regarding the electricity.  
The higher voltage produces a higher current, but the current can be increased also by decreasing the 
resistance.   The quantum self-organization implies that the atomic nucleus finds particular 
configurations which decrease the resistance power.  
%This is similar to the fact that the electric current
%goes through the route with the least resistance, if many circuits are available.

The most favorable configurations and associated SPEs vary for individual eigenstate even within the same 
type of the collective mode.   For instance, %\textcolor{red}{
prolate, oblate or triaxial shapes belong to the quadrupole deformation, but can appear with different patterns of the SPEs within the same nucleus. %}  
The oblate shape is less affected by the quantum self-organization, because smaller numbers of nucleons
on unique-parity orbits are the %\textcolor{red}{
key element %} 
 of the oblate shape in most cases.   In those cases, the organization of
many orbits are rather irrelevant, and the quantum self-organization may not occur to a sizable extent.  This 
feature has been verified with concrete cases.  
On the other hand, many orbits contribute coherently to the prolate deformation, and the quantum self-organization can produce crucial effects.   This has been confirmed by changing the monopole interactions, for instance, closer to the uniform one.   

\begin{wrapfigure} [22]{l}[0cm]{7.3cm}
%\begin{figure} [h]
\begin{center}
\vspace{-0.4cm}
\includegraphics[width=6.8 cm] {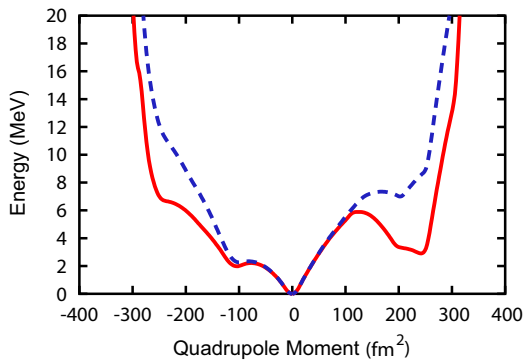}
\caption { %\textcolor{red}{
Potential Energy Surface with the axially symmetric deformation for $^{68}$Ni. 
The red solid line denotes the energy of the constrained Hartree-Fock calculation with the
original Hamiltonian.  The blue dashed line implies the same calculation except that  
the quantum self-organization is suppressed (see the text).
Figure taken from \cite{otsuka2016}.}  %} 
\label{fig:PES}
\end{center}
\end{wrapfigure}

%\textcolor{red}{
We present a concrete example by taking the case of the prolate band in $^{68}$Ni \cite{otsuka2016}.    The monopole interaction between 
the neutron 1g$_{9/2}$ orbit and the proton 1f$_{5/2}$ orbit is more attractive than that between the neutron 1g$_{9/2}$ orbit and the proton 1f$_{7/2}$ orbit mainly due to the robust property of the monopole interaction of the tensor force, and this difference serves as the major origin of the quantum 
self-organization in this particular case: more neutrons in the 1g$_{9/2}$ orbit reduces the 1f$_{7/2}$-1f$_{5/2}$ spin-orbit splitting for protons \cite{otsuka2005,otsuka2010,tsunoda2014}.  The effect  of this difference on the deformation can be seen quantitatively by replacing the strengths of these monopole interactions with the average of their original values, {\it i.e.}, the same value.  Likewise, we reset the monopole interaction between the neutron 1f$_{5/2}$ and the proton 1f$_{7/2}$ orbits and that between the neutron 1f$_{5/2}$ and the proton 1f$_{5/2}$ orbits.   These modifications correspond basically to the removal of the tensor-force monopole contributions, and 
are nothing but the suppression of the present effects of the quantum self-organization.  The resulting Potential Energy Surface is shown in Fig.~\ref{fig:PES} for the axially symmetric deformation compared with that obtained from the original Hamiltonian.  
Around the spherical minimum the energy curves of the two calculations are similar, however, when going to stronger deformation values, the two approaches differ substantially.
%One sees that the resultant energy does not differ between the two calculations around the absolute spherical minimum, but becomes apart from each other for stronger deformation.   
In particular, the prolate profound local minimum, seen in the original calculation (red solid line), is pushed up by about 4 MeV, if the quantum self-organization is suppressed as described above (blue dashed line).  Thus, the quantum self-organization is a part of the crucial mechanisms producing the nuclear deformation.%}

%\end{figure}

 At this point, we mention that the SPE being discussed corresponds somehow to the spherical terms in the Nilsson model \cite{bohr_mottelson_book2} which are comprised of the $\ell \ell$ and $\ell s$ terms as well as the harmonic-oscillator-quanta term.
As their strengths are independent of the deformation, the present effect is not included in the Nilsson model.
 
Type-II shell evolution \cite{otsuka2016} has been discussed, for instance, in Co/Ni region \cite{tsunoda2014,morales2017,leoni2017},    
where neutrons are excited from the $pf$ shell to g$_{9/2}$ across the $N$=40 sub-magic gap.   The neutrons in g$_{9/2}$ and neutron holes in f$_{5/2}$ provide similar sizable monopole effects, as discussed above.  A smaller 
1f$_{7/2}$-1f$_{5/2}$ spin-orbit splitting for protons  
reduces the resistance power against deformation, pulling down the prolate band as seen in Fig.~\ref{fig:PES}.  
Type II shell evolution was introduced as the particle-hole excitation over a magic or sub-magic gap.
Clearly, this kind of mechanism is a very simple and visible case of the quantum self-organization.
On the other hand, the quantum self-organization can occur certainly in more complex ways.   
Such a complex way may be found in the shape transition of Sm isotopes (see Fig.~\ref{fig:2+}), where no magic or sub-magic gap is involved.
We can see the spherical-vibrational-rotational shape evolution in MCSM calculations, as will be 
reported in detail elsewhere.

Likewise, the shape coexistence in Hg/Pb isotopes have been studied.   In those cases, the quantum
self-organization gives intriguing contributions on the pattern of the shape coexistence, as reported
also elsewhere. 

\section{Summary and Perspectives}
We presented a novel mechanism on the relation between single-particle states and collective modes.  A summary is given below.
\begin{itemize}
\item The atomic nuclei are not necessarily like simple rigid vases containing almost free nucleons, 
in contrast to the na\"ive Fermi liquid picture.
\item Nuclear forces are rich enough to change single-particle energies for each eigenstate, leading to 
the quantum self-organization.
\item Single-particle energies can be self-organized, being enhanced by \\
\hspace{0.5cm} (i) two quantum liquids ({\it e.g.}, protons and neutrons)\\
\hspace{0.5cm} (ii) two major force components  \\
\hspace{1.0cm} {\it e.g.}, quadrupole interaction : to drive collective mode  \\
\hspace{1.9cm} monopole interaction \,\,\,\,: to control resistance  
\item Type II shell evolution is a simple visible case involving excitations across (sub)magic gap. 
\item Actual cases such as shape coexistence, quantum phase transition, octupole vibration/deformation, 
super deformation, {\it etc.} can be studied with this scope.
\item The quantum self-organization becomes more important in heavier nuclei where the number of active orbits and 
the number of active nucleons are larger.  With larger numbers of them, the effects of the organization can be more 
significant.  
This feature may be linked to fission and superheavy elements.  On the other hand, the quantum self-organization may not be so visible in light nuclei except for particular cases like intruder bands or cluster (or multiple particle-hole excited) states.  
\item Time-dependent version of quantum self-organization may be of another interest for reactions and 
fission. 
\end{itemize}

\section*{Acknowledgements} 
The presentation and this proceedings are supported by JSPS and FWO under the Japan-Belgium 
Research Cooperative Program.  T.O. thanks Prof. P. Van Duppen for valuable discussions and comments.
The MCSM calculations were performed on K computer at RIKEN AICS (hp140210,hp150224,hp160211,hp170230).
This work has been supported in part by the HPCI Strategic Program (The origin of matter and the universe)
and ``Priority Issue on Post-K computer'' (Elucidation of the Fundamental Laws and Evolution of the Universe)
from MEXT and JICFuS.

\section*{References}
\bibliography{iopart-num}

\end{document}